\documentclass[b5paper,fleqn,usenatbib]{mnras}

\usepackage{newtxtext,newtxmath}
\usepackage[T1]{fontenc}
\usepackage{ae,aecompl}

\usepackage{footnote}
\usepackage{tablefootnote}

\usepackage{graphicx}	
\usepackage{amsmath}	
\usepackage{amssymb}	

\usepackage{verbatim}
\usepackage{bm}
\usepackage{hyperref}

\newcommand{\prv}[1]{\bm{\mathrm{#1}}}
\newcommand{\prm}[1]{\bm{\mathrm{#1}}}

\title[Improved Pulsar Timing]{Improved Pulsar Timing via Principle Component Mode Tracking}

\author[H.-H. Lin et al.]{
Hsiu-Hsien Lin,$^{1}$\thanks{E-mail: hsiuhsil@andrew.cmu.edu}
Kiyoshi Masui,$^{2}$
Ue-Li Pen,$^{3,4,5,6}$
Jeffrey B. Peterson$^{1}$
\\
$^{1}$McWilliams Center for Cosmology, Carnegie Mellon University, Department of Physics, 5000
Forbes Ave, Pittsburgh, PA, 15213,\\ USA\\
$^{2}$Department of Physics and Astronomy, University of British Columbia, 6224 Agricultural Rd,
Vancouver, BC, V6T 1Z1, Canada\\
$^{3}$Canadian Institute of Theoretical Astrophysics, 60 St George St, Toronto, ON M5S 3H8, Canada\\
$^{4}$Canadian Institute for Advanced Research, CIFAR Program in Cosmology and Gravity,
Toronto, ON, M5G 1Z8\\
$^{5}$Dunlap Institute for Astronomy \& Astrophysics, University of Toronto, 50 St George St, Toronto, ON, M5S 3H4, Canada\\
$^{6}$Perimeter Institute for Theoretical Physics, Waterloo, Ontario N2L 2Y5, Canada
}

\date{\today}

\pubyear{2017}

\begin{document}
\label{firstpage}
\pagerange{\pageref{firstpage}--\pageref{lastpage}}
\maketitle


\begin{abstract}

We present a principal component analysis method which tracks and compensates for short-timescale variability in pulsar profiles, with a goal of improving pulsar timing precision.
We couple this with a fast likelihood technique for determining pulse time of arrival, marginalizing over the principal component amplitudes. This allows accurate estimation of timing errors in the presence of pulsar variability.

We apply the algorithm to the slow pulsar PSR J2139+0040 using an archived set of untargeted raster-scan observations at arbitrary epochs across four years, obtaining an improved timing solution. The method permits accurate pulsar timing in data sets with short contiguous on-source observations, opening opportunities for commensality between pulsar timing and mapping surveys.  


\end{abstract}



\section{Introduction}

While there are thought to be 100,000 pulsars in the Milky Way \citep{2004A&A...422..545Y} only about 2600 have been cataloged \citep{2005AJ....129.1993M}. A new generation of drift-scan interferometric telescopes, including CHIME \citep{2014SPIE.9145E..22B}, HIRAX \citep{2016SPIE.9906E..5XN} and HERA \citep{2017PASP..129d5001D}, will soon begin recording data with a primary goal of creating cosmological 21-cm intensity maps.  In addition, there are plans to use MeerKAT \citep{2017MNRAS.466.2780F} to create intensity maps using a raster scan mode.  Such instruments could be used commensally to search for pulsars, and these instruments will have the collecting area, multi-beam capability, and on-sky integration time to substantially increase the pulsar discovery rate. 

The pulsar search will likely have to work within the observational parameters of the hydrogen surveys. For the drift-scan telescopes, the duration of a contiguous observation will be limited to the time sources take to drift through a beam, which for CHIME will be a few minutes. MeerKAT, with its raster scan observing mode will make shorter duration passes but will pass near a pulsar's coordinates more often. Integration time can be accumulated by combining multiple passes. Such a coherent pulsar search in time-disjointed data has been demonstrated in \citet[Chapter 4]{Anderson:1993cna}. 

To further develop and test time-disjoint pulsar search algorithms, we used an existing hydrogen-mapping data set, the Green Bank Hydrogen Intensity Mapping (GBTIM) survey \citep{2010Natur.466..463C, 2013ApJ...763L..20M, 2013MNRAS.434L..46S}. This is the data set in which fast radio burst FRB~110523 was discovered \citep{FRB110523}. The set consist of raster scans, so it contains a time-disjointed set of near-passes of pulsar coordinates.

Our long-term goal is to search for new pulsars, but first we need to
test our algorithms and code on known examples, so we obtained a timing solution for PSR~J2139+0040.
The pulsar was discovered by \citet{1996ApJ...469..819C} at right ascension 21:39:42(16), declination +00:36(5), period 0.312470(3)\,s, and dispersion measure of 36(7)\,pc/cm$^3$. The pulsar is bright---we observe an average flux at 800\,MHz of approximately 50\,mJy---such that single pulses are readily detectable with the Green Bank Telescope. Our team began studying this pulsar because it lies within two degrees of FRB~110523 and could be used to constrain the Galactic component of the scintillation properties for this region of the sky.

Despite the sporadic scan-mode observations in the data set, we show that a timing solution for PSR~J2139+0040 can be obtained, and we present improved timing parameters. We find timing residuals are reduced by use of a novel principal component analysis (PCA) technique to fit the time-variable pulse waveforms. The PCA technique automatically compensates for the rapid mode switching of the pulse shape, allowing a more precise timing solution. We describe this technique in detail since it may be more widely useful in pulsar timing.

\section{Data and Processing}

Here we describe our data and observations as well as the processing of the data into folded pulse profiles.

\subsection{Observations}
The data set we used came from an 21-cm intensity mapping survey, in which we raster-scanned four different Wiggle-Z (WZ) fields (1hr, 11hr, 15hr, and 22hr)\citep{2007ASPC..379...72G} recording spectra from 700-900 MHz with 4096 spectral frequency channels, using integration intervals of 1.024\,ms. The full survey is comprised of 680 hours of observations between 2011 and 2015. PSR~J2139+0040 is located in the 22hr field, which was observed as part of GBT projects 10B-336 and 14B-339. Scans of this field have the beam crossing close to the pulsar often, but for short periods of time, with the beam typically moving across the sky at a rate of several degrees per minute. The set consists of sporadically-spaced observing sessions of several hours. This provides a sample of pulsar data with a wide range of crossing angles and with baseline drifts in the data due to variable ground spill and sky brightness structure. The cadence of the observing epochs is variable. It is therefore challenging to extract a precise pulsar timing solution from the set, providing a test of our processing techniques.

Our WiggleZ data are stored in blocks of 2048 time samples ({\tt PSRFITS} records \citep{2004PASA...21..302H}). From our bulk data, we select all such blocks where at the block midpoint the telescope boresight is within 0.15$^{\circ}$ from the published pulsar position. This corresponds roughly to a half width at half maximum of the telescope beam at 800\,MHz. The resulting data set includes 1975 seconds of integration time, which we term the WZ data set.

In addition to the raster scan data described above, we obtained a single pointed observation of PSR~J2139+0040 with a duration of one hour on MJD 57178. The frequency range of the pointed data is 720-920 MHz with 2048 spectral frequency channels, and an integration interval 8.192e-5\,s, grouped into records of 4096 samples. We dub this the pointed data set.

Using the {\tt PRESTO}\footnote{\url{http://www.cv.nrao.edu/~sransom/presto/}} software package on the pointed data set yields DM = 31.7262 pc cm$^{-3}$, which we use for all subsequent analysis except Section~\ref{subsection: Dispersion Measure} where we perform a full fit for the DM.

\subsection{Preprocessing}
\label{subsection: preprocessing}

Data pre-processing produces an initial calibration of the data, removes flux from  a noise injection source, mitigates radio-frequency interference (RFI), long-time-scale noise, and other sky signals such as point sources moving through the beam. 
To remove the system bandpass response function, we divided the recorded intensity by its time mean, converting to units of the system temperature.

To mitigate RFI and long-time-scale signals we apply a stack of filters to the data consisting of: 1.~deleting frequencies where the time variance is anomalously high, 2.~subtracting the time mean from the data, 3.~subtracting the time-linear component from the data, 4.~deleting time samples where the frequency mean is outlying, 5.~subtracting the frequency mean from the data, and finally repeating step 1. We use $5\sigma$ for all thresholds.

The above filters result in a zero-mean artifact in the folded profiles where the pulse profile is pushed negative just outside the main pulse. This is visible in the middle panel of Figure~\ref{figure: folding_dedisperse_fit}. However, this artifact is a result of linear operations on the data and as such does not bias our subsequent timing analysis.

\subsection{Barycentric Time and Folding}

We choose to time the pulsar in the Barycentric coordinate system because we are working toward searching such data sets for weaker pulsars. This step is not critical to the mode-tracking technique presented in Section \ref{subsection: Fitting phase bin number by SVD analysis}, but we describe it for completeness. We converted time stamps from Universal Time to Barycentric Time, $\tau(t, \alpha, \delta)$, defined as the time a signal from a far-off source at right ascension $\alpha$ and declination $\delta$ arriving at the observatory (here GBT) at time $t$, would arrive at the solar system barycenter.

We perform this conversion using the {\tt TEMPO}\footnote{\url{http://tempo.sourceforge.net/}} software package as invoked by the {\tt bary} command in {\tt PRESTO}. For the right ascension and declination we initially use the previously published location of the pulsar, which we later refine. As described in Section~\ref{subsection: Fitting a timing solution}, refining this position requires the partial derivatives of the Barycentric times with respect to right ascension and declination, which we calculate by finite difference.

To form pulse profiles, we group our sporadically spaced WZ data into groups spanning five minutes or less. These are typically very sparsely sampled with most groups containing only a few seconds of on-target data. The five-minute duration was selected to reduce the set to a manageable size of fixed cadence, with reasonable signal-to-noise for each folded waveform, and to ensure little worry of phase drift within a group.
The pointed data, in contrast, is contiguously sampled and we divide it into 640 groups of duration 5.369\,s, which we use to characterize the time variability, and later stack into 40 pulse profiles of duration 86\,s used for timing.

We fold the time stream on the pulsar period 0.312470\,s \citep{1996ApJ...469..819C}.
We use 200 phase bins for the WZ data, and 800 phase bins for pointed data.

\begin{figure}
\centering
\includegraphics[width=\columnwidth]{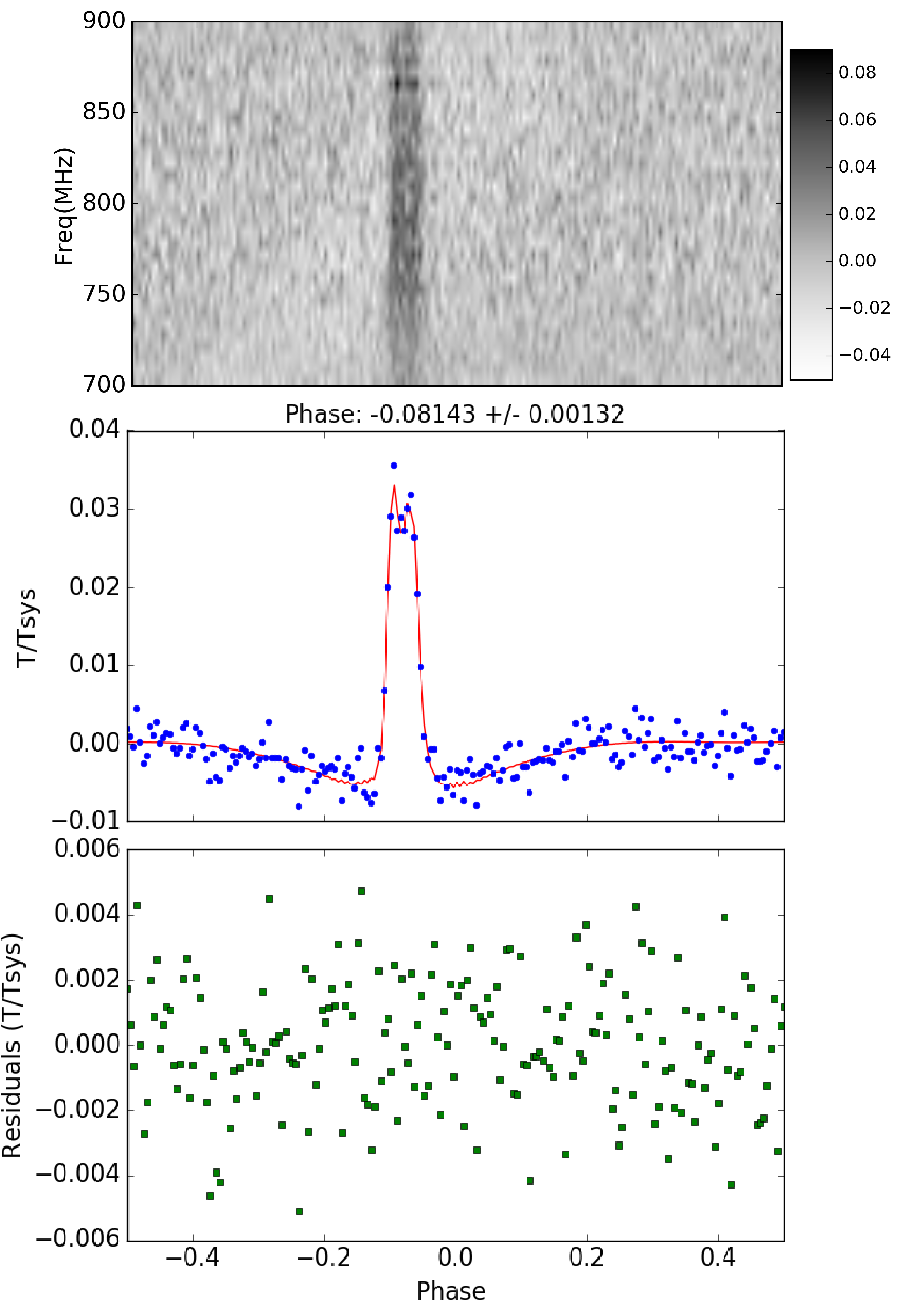}
\caption{{\bf Typical Pulse profile for PSR J2139+0040.}  Upper panel: Dedispersed dynamic spectra. Middle panel: Blue dots are the pulse profile averaged across frequency, and red line is the best-fit model.  Barycentric Arrival Times are used in these plots. Bottom panel: Fit residuals.}
\label{figure: folding_dedisperse_fit}
\end{figure}

We dedisperse the folded data, and then average it over frequency yielding pulse profiles as shown in Figure~\ref{figure: folding_dedisperse_fit}. This yields 289 such profiles although a number of these are noise dominated and show no evidence of the pulsar. We discard these, resulting in 232 pulse profiles in the WZ data set. Combined with the 40 profiles from the pointed data set, there are a total of 272 pulse profiles used for the timing analysis. Figure~\ref{figure: J2139_all} below shows how these profiles are distributed in time.

\section{Analysis and Results}

Here we describe the analysis of the folded pulse profiles. Our analysis accounts for the pulsar's short-time-scale variability, estimates the pulsar phase, and fits a timing solution.

\subsection{Principal component analysis of pulse shapes}\label{subsection: Fitting phase bin number by SVD analysis}

\begin{figure}
\centering
\includegraphics[width=\columnwidth]{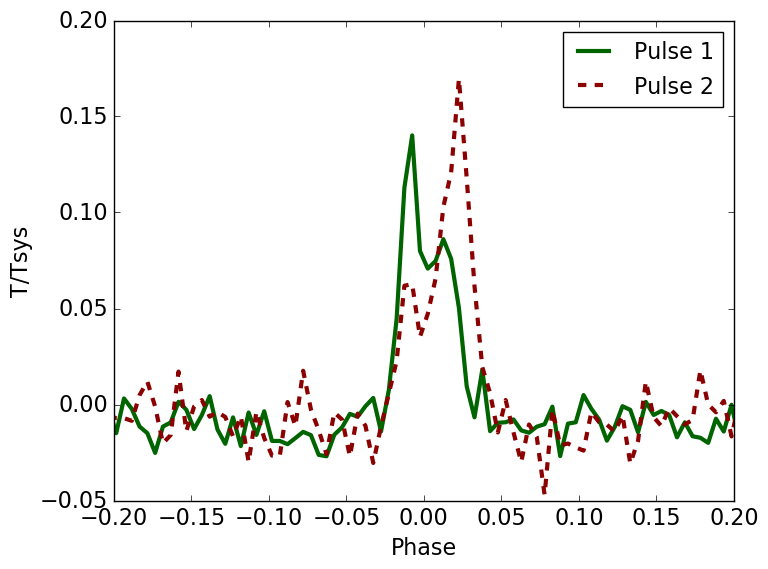}
\caption{{\bf Pulse shape variations.} The two waveforms show folded data from the WZ set, focusing on the  phase range -0.2 to 0.2. The pulsar shows mode-switching behavior: The ratio of the fluence between the first and second peak varies substantially. If these two waveforms were naively fit to an average profile, there would be a substantial timing error because of this profile variation.}
\label{figure: two_pulses}
\end{figure}

Examining the set of folded pulse profiles, we find substantial variation in shape. In particular, some profiles have a stronger first peak, while others have a stronger second peak as shown in Figure~\ref{figure: two_pulses}.
This is the dominant mode of time variability for the profiles, with the preference for one peak over the other appearing to be correlated over timescales of several minutes.
This is thus a form of mode switching, which is common for a substantial fraction of pulsars on a range of time scales {\citep{2004hpa..book.....L}}.
To find the best-fit arrival time for each folded waveform, one could simply fit each waveform to an average profile, but waveform shape variations would then cause timing errors, since the fit would favor earlier arrival times when the first peak is stronger. To address this, we use a principal component analysis of the pulse profiles (PCA), which automatically tracks the pulsar mode. This PCA Mode Tracking compensates for mode-switching waveform-shape variation.

\begin{figure}
\centering
\includegraphics[width=\columnwidth]{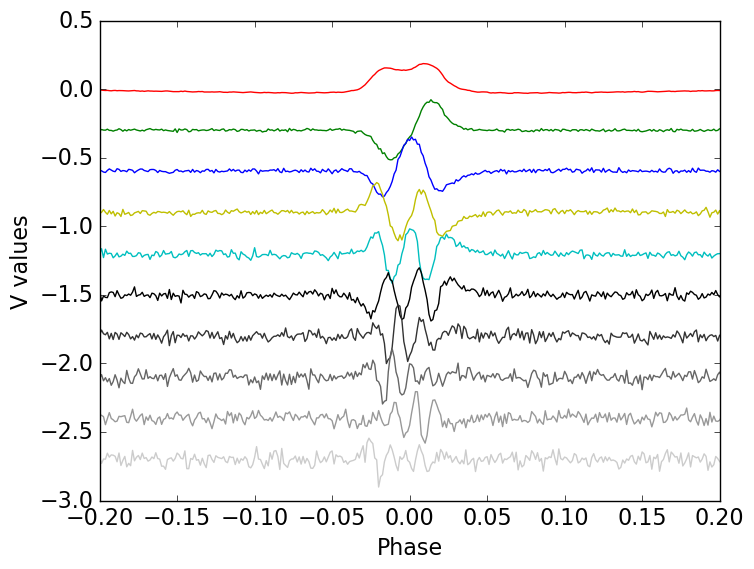}
\caption{The 1st to the 10th modes of $\prm V$ are in the sequence of red, green, blue, yellow, cyan, black and dark to light greys, respectively. We set the offset of the 1st to the 10th modes by steps of -0.3.}
\label{figure: v_mode}
\end{figure}

We use the 640 pulse profiles from the pointed data (each containing 5.369\,s of integration) to construct the PCA. Using Fourier techniques to achieve sub-bin alignment, we align these profiles according to a preliminary, linear timing solution valid for the pointed data. We create a matrix of 640 pulse profiles called $d_{E i}$ where $E$ is the epoch, an index of the mid-point time
of each folding interval, and $i$ is the 800-point phase bin within the waveform. We carry out a Singular Value Decomposition:
\begin{equation}\label{equation: svd}
\centering
d_{Ei} =\sum_m {U_{Em}}{S_{mm}}{V_{mi}},
\end{equation} 
where for each mode $m$: $U_{Em}$ is the eigenfunction in epoch $E$, $V_{mi}$ the eigenfunction in phase $i$, and $S_{mm}$ is the singular value. We show $V_{mi}$ for the first ten $m$ modes in Figure~\ref{figure: v_mode}. 

Inspecting the phase eigenfunctions, the first mode $V_{0 i}$ appears similar to the average waveform. We tested this equivalence by averaging the 640 pulse profiles into single template. We compared this average template with the mode $V_{0 i}$ of our PCA technique, and found the two consistent within the noise.

The second mode $V_{1i}$ allows the two peaks in the waveform to depart in amplitude from the average. Essentially, $U_{E1}$ tracks the variation with epoch of the waveform mode, changing sign when the mode switches.

Since we intend to use the $V$ modes as templates for fitting pulsar phases, a concern is that noise in these modes will bias the phase measurements. To reduce this noise for the modes above the first, we set them to zero outside of the main pulse (all phase bins except the central 160 of 800) since we see no evidence for pulsar flux outside this region. For the primary mode, the filtering artifact mentioned in Section~\ref{subsection: preprocessing} results in a smooth structure outside this region and so we spline the profile outside the central 160 phase bins.

\subsection{Pulsar time of arrival estimation from direct integration of the likelihood}\label{subsection: Pulsar time of arrival estimation from direct integration of the likelihood}
The natural next step would be to simultaneously fit mode amplitudes and a pulse phase to each of our individual pulse profiles, using $V_{mi}$ as templates. However, we find that for this high-dimensional parameter space, the likelihood often has multiple maxima, and so the traditional least-squares fitting method fails. The multi-modal nature of the parameter space is due to the fact that for some profiles, there are multiple combinations of the $V_{mi}$ modes that, with different phases, may adequately describe the profiles such as those shown in Figure~\ref{figure: two_pulses}. To deal with this, we employ a new pulse profile fitting technique that fully samples the parameter space.

We denote an individual measured folded pulsar profile $d_i$ where the index $i$ runs over the phase bins (we suppress the index $E$ for epoch in this section, since the analysis is performed independently for each profile). We model the measured profile as
\begin{equation}
    d_i = \sum_m A_m V_{mi}(\phi) + n_i,
\end{equation}
where $V_{mi }(\phi)$ are the profile template modes (with $m$ running over mode number), $A_m$ is the mode amplitude, $\phi$ is the finely adjusted pulsar phase, and $n_i$ is the noise contribution. The dependence of $V_{mi }$ on $\phi$ describes the rotation of the templates to match the data. For notational brevity, we now switch to vector notation, where the above equation becomes:
\begin{equation}
    \prv d = \prv A^T \prm V(\phi) + \prv n.
\end{equation}
We assume Gaussian noise that is uniform and uncorrelated: $\langle n_i n_j \rangle = \delta_{ij} \sigma^2$.
The template modes have already been measured so the parameters of the model are $\phi$ and the amplitudes, $\prv A$.

Of these parameters we are chiefly interested in $\phi$. All the information about this parameter is in $p(\phi|\prv d)$, the posterior probability distribution of $\phi$ given the measurements $\prv d$. For instance, the measurement mean and variance of the phase are the first and second moments of this distribution. This can be calculated from the posterior of all the parameters marginalized over the mode amplitudes:
\begin{equation}
    \label{e:margin_post}
    p(\phi|\prv d) = \int d^N{\prv A}\, p(\phi,{\prv A}|\prv d),
\end{equation}
where $N$ is the length of the vector $\prv A$.

For flat priors on the initial parameters, Bayes' Theorem states that the posterior distribution for the parameters is proportional to the likelihood function, $p(\phi,{\prv A}|\prv d) \propto p(\prv d | \phi,{\prv A})$. Since the data is Gaussian-distributed, the latter is given by:
\begin{align}
    p(\prv d | \phi,{\prv A}) &\propto
     \exp\left[
        {-\frac{1}{2}\chi^2(\prv d, \phi,{\prv A})}
    \right]
    \\
    \chi^2(\prv d, \phi,{\prv A}) &= 
    \frac{1}{\sigma^2}\sum_i \left[
        d_i - \sum_m  A_m V_{mi}(\phi)
    \right]^2.
\end{align}

Combining the above, we have:
\begin{equation}
    p(\phi|\prv d) \propto \int d^N{\prv A}\,\exp\left\{-\frac{1}{2\sigma^2} \sum_i \left[ d_i - \sum_m A_m  V_{mi}(\phi) \right]^2 \right\}
\end{equation}
This is a multidimensional integral over the vector space of $\prv A$, which would be prohibitively expensive to do numerically. A key insight is that since the model is linear in $\prv A$, the likelihood is Gaussian not only in the data, but in $\prv A$ as well (but not in $\phi$). This permits the integral to be done analytically.
This same insight was used in \citet[Chapter 2]{2013PhDT..Pennucci} where a similar integral appears over the frequency dependant template amplitudes in wide-band pulsar timing. However, while in wide-band timing this is a computational convenience, here being able to perform this integral is essential due to the multi-modal nature of the likelihood space. Similarly, the integral appears in \citet{2015MNRAS.447.2159L} when marginalizing over epoch dependant profile amplitudes in profile domain timing analysis.

When performing the integral, there is a factor that depends on the expression
$$
    \sum_i V_{mi}(\phi) V_{ni}(\phi).
$$
This expression is independent of $\phi$ because the dependence of $V_{mi}$ on $\phi$ is just a shift of the template modes, which disappears after summing over phase bins. Thus, there is no need to carry out this part of the calculation. Note that this is only true if the noise is uniform.
The final expression is
\begin{align}
    p(\phi|\prv d) 
    &\propto
    \exp\left[
        -\frac{1}{2\sigma^2} 
        \sum_m \sum_i V_{mi}(\phi)  d_i \sum_j V_{mj}(\phi) d_j,
    \right]
\end{align}
or equivalently
\begin{align}
    p(\phi|\prv d)
    &\propto
    \exp\left[
        -\frac{1}{2\sigma^2} 
        (\prm V \prv d)^T
        \prm V \prv d
    \right].
\end{align}

which is proportional to
\begin{align}
    p(\phi|\prv d)
    &\propto
    \exp\left\{
        {-\frac{1}{2}\chi^2\left[\prv d, \phi,\hat{\prv A}(\prv d, \phi)\right]}
    \right\},
\end{align}
where $\hat{\prv A}(\prv d, \phi)$ is the linear-best-fit value for the template amplitudes at fixed $\phi$. Linear regression gives
\begin{align}
    \hat{\prv A}(\prv d, \phi) = \left[\prm V(\phi) \prm V^T(\phi)\right]^{-1}\prm V(\phi) \prv d.
\end{align}

\begin{figure}
\centering
\includegraphics[width=\columnwidth]{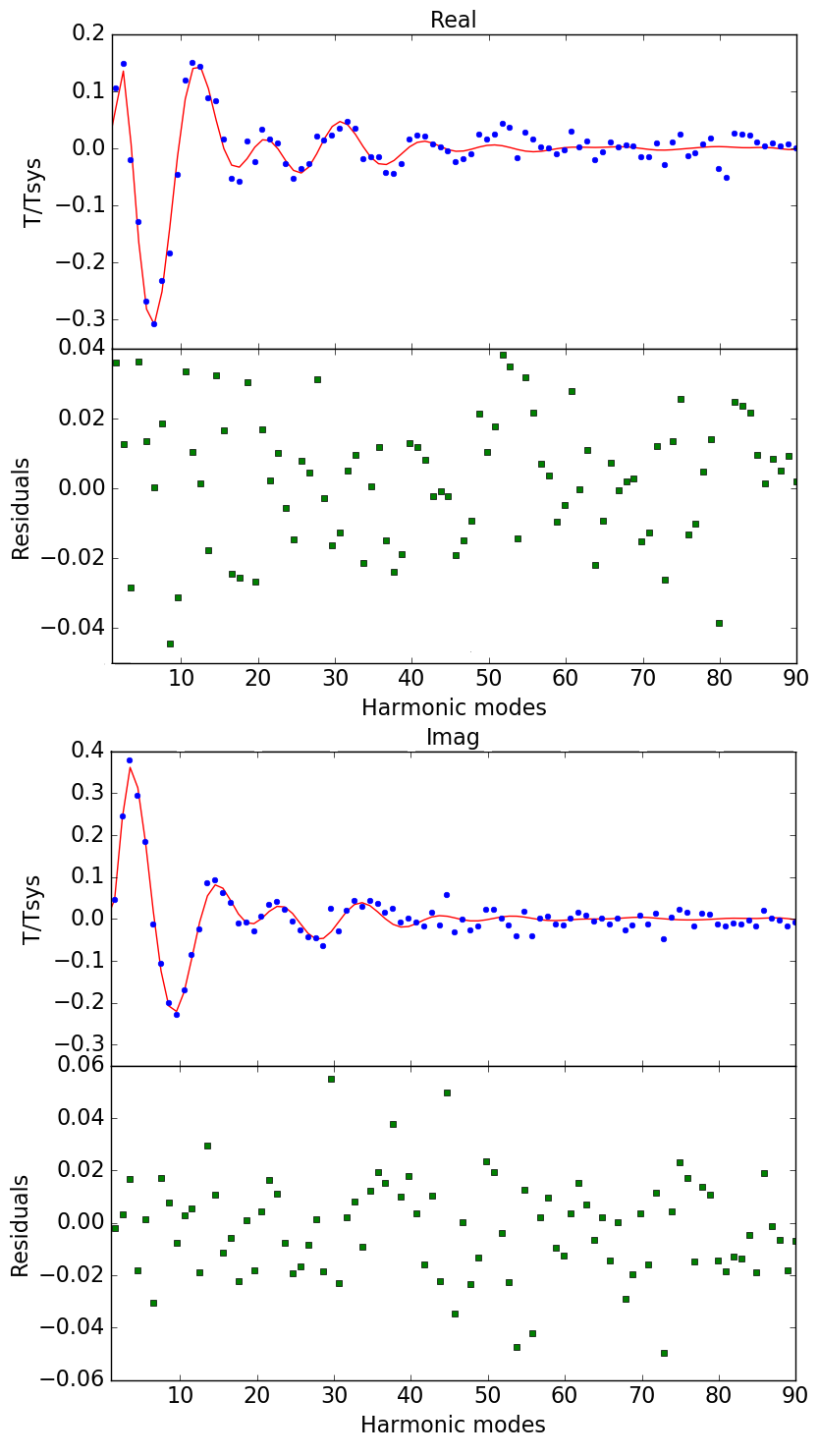}
\caption{{\bf Pulse profile fit residuals in the frequency domain.}  Real and imaginary parts of the Fourier components are shown. The red curve and blue dots are the model and the measured flux, respectively. Green squares in lower panel are residuals. Fitting for phase in the frequency domain allows for simple interpolation with time precision finer than the 1.024 ms cadence of the dynamic spectra.
}
\label{figure: fft_Fourier}
\end{figure}

So succinctly, the procedure for estimating the time of arrival is:
\begin{enumerate}
    \item At fixed $\phi$ perform a \emph{linear} fit for the mode amplitudes
    $\prv A$, which we found in Section~\ref{subsection: Fitting phase bin number by SVD analysis}.
    \item Evaluate $\chi^2$ for these parameters and take $e^{-\chi^2/2}$ as the likelihood. \label{item: evaluate the likelihood}
    \item Repeat for a range of $\phi$ values, covering the region of low $\chi^2$ that dominates the likelihood. (We repeat this step for a range of -20 to +20 phase bins from the fixed $\phi$, with steps of 0.02 phase bins.)
    \item Take the zeroth, first, and second moments to get the normalization, phase estimate, and variance, respectively. In Section~\ref{subsection: Fitting a timing solution}, we use the phase estimate and variance to find the timing solution.
\end{enumerate}

While our procedure has been described in the profile space, our actual analysis is performed in the Fourier domain, as in \citet{1992RSPTA.341..117T}. We use only the first through 90\textsuperscript{th} harmonics to limit contamination from noise in the template modes and observed extraneous noise near 300\,Hz. We use $N=6$ template modes in our fits. An example of a profile fit is shown in Figure~\ref{figure: fft_Fourier}.

Finally, properly estimating the error on $\phi$ requires an estimate of the noise power $\sigma^2$. We use the value of $\sigma^2$ that results in a reduced chi-squared of unity ($\chi^2/{\rm \nu}$, where $\nu$ is the number of degrees of freedom) for the best fit (maximum likelihood) parameters.

\subsection{Timing solution}
\label{subsection: Fitting a timing solution}

We proceed to adjust the parameters of the pulsar timing model:
\begin{equation}
\phi_E =
{\phi_0}
+ \frac{\Delta P}{P^2}(\tau_E - \tau_0)
+ \frac{\dot P}{2 P^2}(\tau_{E} - \tau_0)^{2}
+ {\frac{1}{P}}\left[-\frac{\partial \tau_E}{\partial \alpha} \Delta{\alpha}
      - \frac{\partial \tau_E}{\partial \delta} \Delta{\delta}\right]
\label{equation: timing_model}
\end{equation}
where, $\phi_E$ is the barycentric phase of the pulse profile indexed by $E$, $\tau_{E}$ is its barycentric time, $\tau_{0}$ is the reference epoch, $P$ is the period of the pulsar used for folding, ${\partial \tau_E}/{\partial \alpha} $ is the derivative of the barycentric time with respect to source right ascension, and ${\partial \tau_E}/{\partial \delta}$ is derivative of the barycentric time with respect to source declination. There are also five free parameters in the timing model, including the period derivative $\dot P$, period correction $\Delta P$, initial phase offset $\phi_0$, right ascension correction $\Delta{\alpha}$, and declination correction $\Delta{\delta}$. We proceed to adjust these parameters such that the above equation fits our phase measurements using standard weighted least squares and extract best fit parameters with uncertainties.

One challenge to obtaining a timing solution with this data set is the large gap in our data. We have phase measurements for epochs over several months in 2011 and 2015 but none in between. Compared to more uniformly sampled data, this distribution of measurements weakly constrains the total number of pulsar rotations in the gap period. Indeed we find multiple $\chi^2$ minima corresponding to changing the parameter $\Delta P$ by integer multiples of $\sim 8\times 10^{-10}$\,s, which changes the number of rotations in the gap. However, the lowest of these minima is $\delta \chi^2=44.5$ smaller than the next lowest. Thus this minimum is strongly preferred over the others, and our timing parameters are well constrained.

Inspecting the timing residuals, we find one phase measurement that is an extreme, $9\sigma$, outlier. Inspecting the full likelihood curve for the profile phase fit (as described in Section~\ref{subsection: Pulsar time of arrival estimation from direct integration of the likelihood}), we find the likelihood to be very non-Guassian, making such an outlier roughly $0.1\%$ likely. This is not terribly improbable given that we have 272 such data points, and is far more likely than a $9\sigma$ outlier for Gaussian data. Since our least-squares fit to the timing solution implicitly assumes Gaussian-distributed phase errors, we exclude this outlying data point in the timing solution fit.

\begin{figure*}
\centering
\includegraphics[width=\textwidth]{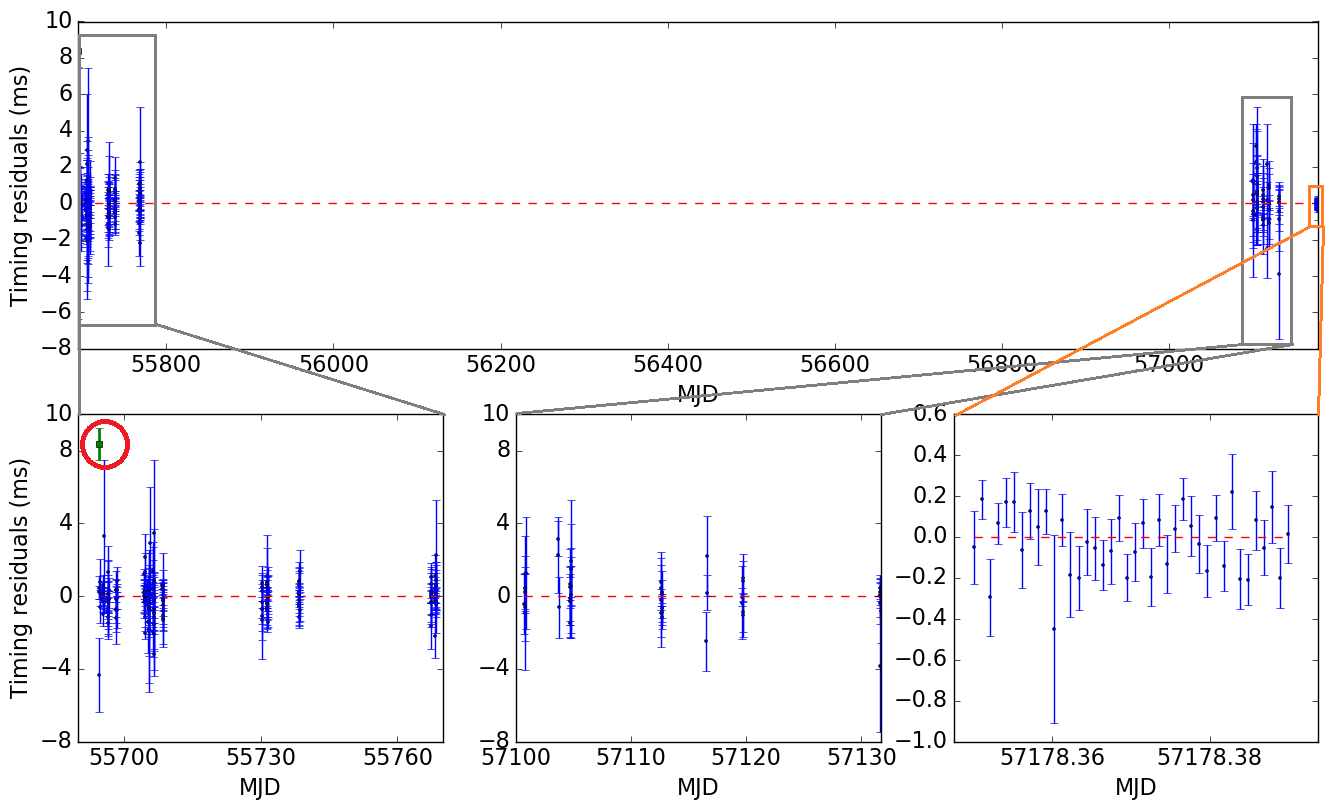}
\caption{\textbf{Timing residuals with error bars of WZ data from MJD 55694 to 57178.} The upper panel shows the residuals over five years. The left lower panel shows detail of the residuals of 2011 WZ data. The denoted with a circle around in the left lower panel is an outlier resulting from the non-Gaussian nature of the error bars, which we exclude from the timing solution. The lower middle panel shows the 2015 WZ data. The lower right panel shows the 2015 pointed data. All data are shown offset from the best fit model indicated with the dashed red line.
}
\label{figure: J2139_all}
\end{figure*}

The timing residuals are shown with the fitting results in Figure~\ref{figure: J2139_all}.
The fit has a reduced chi-squared of $\chi^2/{\nu}=1.26$ for 266 degrees of freedom, which is reasonable given the non-Gaussian nature our phase measurements which we have represented with 1-$\sigma$ error bars. We inflate the uncertainties on the timing parameters by this factor.

The parameters for the final timing solution for PSR~J2139+0040 are given in Table~\ref{table: J2139+00}.  We find new right ascension and declination, period, period derivative, reference time of arrival (derived from $\phi_0$), and also provide the epoch used for the timing model. We use the period and period derivative to estimate the magnetic field strength ($B$), the characteristic age ($t_c$), and the spin-down luminosity ($\dot E$)\citep{2004hpa..book.....L}.

\begin{table*}
\centering
\begin{tabular}{ |l|l| } 
 \hline
Right ascension, $\alpha$                           & $+21:39:25.20(2)$\\
Declination, $\delta$                               & $+00^{\circ}40^{'}21.6(3)^{"}$\\
Period, $P$ (s)                                     & 0.3124695464326(2)\\         
Period derivative, $\dot P$ (s s$^{-1}$)            & $7.64(13)\times 10^{-18}$\\            
Dispersion measure, DM (pc\,cm$^{-3}$)              & 31.585(8) \\
Period epoch, $\tau_0$ (MJD)                        & 56436.5 \\
Reference arrival time (Barycentric MJD, 900\,MHz)  & 56436.506782091(13) \\
Magnetic field strength, $\log_{10}[ B/{\rm G}]$    & 10.7 \\
Characteristic age, $\log_{10}[ t_c/{\rm yr}]$      & 8.8 \\
Spin-down luminosity, $\log_{10}[ \dot E / ({\rm erg}/{\rm s})]$    & 31.0 \\
Dispersion-derived distance, $d$ (kpc)              & 1.7 \\
 \hline
\end{tabular}
\caption[]{
\textbf{Timing solution and measured parameters of PSR J2139+0040.} 
}
\label{table: J2139+00}
\end{table*}

To check for evidence of proper motion of the pulsar's sky position, we fix all parameters other than the position, then separately fit the 2011 and 2015 data. We find the two output positions are consistent with each other and the solution using the full data set, and as such we find no evidence for proper motion.

We used the {\tt TEMPO} pulsar timing software to verify our timing solution. For this we converted our pairs of epoch-phase measurements to barycentric times of arrival then inverted {\tt PRESTO}'s {\tt bary} command to convert to topocentric times of arrival. Feeding these and our timing solution parameters into {\tt TEMPO} we find agreement in the value of $\chi^2$. Letting {\tt TEMPO} refit our timing-model parameters we find no significant shifts.

\subsection{Dispersion Measure}
\label{subsection: Dispersion Measure}
We used the pointed data to determine DM. We align the 40 folded profiles in phase based on our timing solution, then stacked them into a single profile. 
We fit this profile for the dispersion using the first $V$ mode as a template according to: 
\begin{align}
d_{if} &= A\left({\frac{f}{800\,{\rm MHz}}}\right)^{\beta}V_{0i}(\phi_f)\label{equation: dm_solution}\\ 
\phi_f &\equiv \phi_{900} + \frac{\rm DM}{P}
\frac{4148.808\,{\rm s\,MHz^2}}{\rm pc / cm^3}
\left[\frac{1}{f^{2}}-\frac{1}{(900 {\rm MHz})^{2}}\right]
\label{equation: dm_phase}
\end{align}
Here, $d_{if}$ is the frequency-dependant profile data, which are fit with parameters for the amplitude $A$, power-law slope $\beta$, and frequency dependant phase $\phi_{f}$ for each frequency $f$. The frequency dependant phase $\phi_{f}$ contains an offset $\phi_{900}$ and the dispersion phase delay.

 We subsequently calculate the dispersion-measure-based distance using Cordes-Lazio NE2001 Galactic Free Electron Density Model \citep{2002astro.ph..7156C} website\footnote{\url{https://www.nrl.navy.mil/rsd/RORF/ne2001}} using the new right ascension, declination, and DM.
 The results of this analysis are given in Table~\ref{table: J2139+00}

\subsection{Timing improvement available from PCA Mode Tracking}
\label{subsection: timing impovement}

To assess the impact of PCA mode tracking we compared the timing residuals using long and short integrations. We carried out these parallel analyses on sets of 16 individual 5.4\,s profiles from the pointed data. First, we stacked these profiles into a single pulse profile with a total integration time of 86\,s and fit this profile with the technique described in Section~\ref{subsection: Pulsar time of arrival estimation from direct integration of the likelihood}. Second, we fit the 16 profiles individually, then combined their resulting posterior distributions. We did this for the 40 sets of 16 profiles covering the full hour of pointed data. The first technique is similar to the conventional technique of taking long averages to smooth out pulse-to-pulse variability. The second technique uses fine grained data allowing the PCA algorithm to track and compensate for the variability.

We find that using PCA mode tracking and fitting fine-grained 5.4 second averages results in a $\sim20\%$ smaller uncertainty on average in the pulse time of arrival (calculated from the second moment of the posterior as described in Section~\ref{subsection: Pulsar time of arrival estimation from direct integration of the likelihood}). While the improvement is substantial for this pulsar, it may not be generic. Pulsars are highly individual: they vary substantially in the degree and time scales mode switching, so we anticipate the improvement available from PCA mode tracking will also vary strongly from case to case.

\section{Discussion and Conclusions}

Pulsar timing observations often make use of profiles averaged for minutes. These long averages smooth out the effect of rapid mode switching. Since it is in general sub-optimal to average over a time-variable signal, we suggest that PCA Mode Tracking could be applied broadly, by intentionally using short averages that display rapid mode switching---which the PCA then follows and compensates for.  
In particular, it may be useful to test this technique using data from millisecond pulsars, since improvement of timing precision of these pulsars allows more precise tests of general relativity, along with tighter constraints on long-wavelength gravitational wave backgrounds. 

\begin{figure}
\centering
\includegraphics[width=\columnwidth]{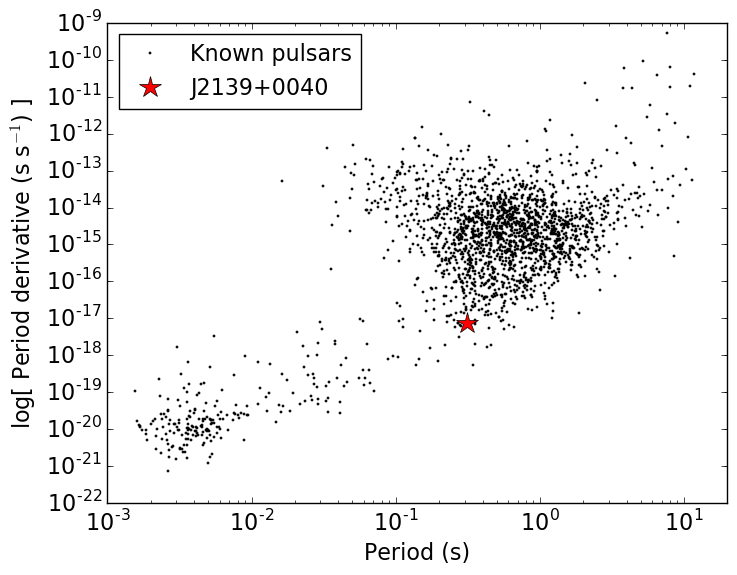}
\caption{Periods and Period derivatives of known pulsars\citep{2005AJ....129.1993M}. Pulsar~J2139+0040 lies at low $\dot P$ compared to other pulsars of similar period, but is not an extreme outlier.}
\label{figure: psr_p0_p1}
\end{figure}

We have obtained a substantially improved timing solution for PSR~J2139+0040, using untargeted observations in conjunction with a $\sim$1 hour duration targeted observation. Our data are well-described by a simple five-parameter model,  including its sky location, reference phase, period, and period derivative. The measured slow-down rate of $\dot P = (7.64\pm0.13)\times10^{-18}$(s\,s$^{-1}$) is well below average, with $\sim 99\%$ of slow pulsars having a higher rate (\citet{2017MNRAS.467.3493J}) as shown in Figure~\ref{figure: psr_p0_p1}.  This indicates that PSR J2139+0040 has an abnormally weak magnetic field.

For many of our pulse profiles, we find a highly non-Gaussian likelihood with multiple maxima for the phase. However, by analytically integrating over the template amplitudes, we are able to fully sample the marginalized posterior for phase and calculate the distribution's mean and standard deviation. The subsequent fit to a timing model achieves a reduced chi-squared of 1.26, which is reasonable given the non-Gaussian nature of our phase errors. A more complete treatment would be to use the phase measurement posteriors directly when fitting a timing solution, rather than reducing them to their first and second moments. This would optimally extract the timing information from the phase measurements but would substantially complicate fitting a timing solution, since least-squares would no longer be applicable.

Our PCA mode tracking method addresses issues similar to those addressed via  "profile-domain pulsar timing" \citep{2015MNRAS.447.2159L, 2015MNRAS.454.1058L, 2017MNRAS.466.3706L}, particularly, what these authors refer to as ``low-frequency stochasticity'' and ``phase-correlated stochasticity''. The profile-domain techniques uses pre-defined uncorrelated shapelet components to contribute to the individual profiles.
In contrast, the PCA mode tracking method is non-parametric. There is no need to guess the shaplets, the PCA finds them automatically. Compared to the profile-domain strategy we anticipate that fewer components will be needed, yielding less degeneracy with the pulse phase and likely reducing timing error.  The PCA mode tracking technique derives its mode templates from the data, and so for weak pulsars these waveforms may have noise that is absent when using the pre-defined templates of the profile domain technique. 

A PCA technique was employed to characterize pulsar variability in \citet{2011MNRAS.418.1258O}, where, rather than fitting for the time of arrival and component amplitudes in high cadence data, the authors used the empirical correlation between the timing residuals and component amplitudes to retroactively correct for systematic timing error. 

Our results demonstrate the feasibility of timing pulsars commensally with mapping surveys. In particular, upcoming hydrogen intensity mapping surveys using MeerKAT \citep{2017MNRAS.466.2780F} and Phase 1 of the Square Kilometre Array \citep{2015aska.confE..19S} will map large fractions of the southern sky with thousands of hours of telescope time. The telescopes will operate as a collection of single dishes and will thus need to scan rapidly to overcome $1/f$ noise. There is the potential to obtain pulsar timing solutions for free (in terms of telescope time) using the data from these surveys in much the same way as we have done here. The challenge is in storing the data at the rapid cadence required for pulsar studies, but, in single dish mode, data volumes are modest compared to interferometric mode. Mapping surveys spend a small fraction of their time pointing at known pulsars, however the mapping survey gets this data at many epochs, for every pulsar in its survey field, providing a large volume of timing data.

One of our goals in this work is to enable pulsar \emph{searches} commensal with intensity mapping experiments. Having obtained a precise timing solution for a known pulsar in the GBTIM data, we have demonstrated that accumulation of pulsar data over five years in few-pulse snippets can be accomplished, although this may be more difficult for weaker or more erratic pulsars. 

This analysis was possible because the GBTIM data used short 1 ms integrations. The intensity mapping data from CHIME \citep{2014SPIE.9145E..22B} and HIRAX \citep{2016SPIE.9906E..5XN} will use longer integrations for their intensity mapping data, but these instrument will have additional transient-search backend hardware allowing high cadence analysis of the data streams. New search algorithms have recently been proposed by \citet{2016arXiv161006831S} to reduce the cost of such long-term pulsar waveform assembly by several orders of magnitude. If these techniques could be employed at upcoming transit survey instruments such as CHIME and HIRAX the pay off could be a substantially increased rate of pulsar discovery. 

\section*{Acknowledgements}

We thank Kendrick Smith, Ingrid Stairs, Maura McLaughlin and Alexander Roman for valuable discussions.
K.~W.~M.~is supported by the Canadian Institute for Theoretical Astrophysics National Fellows program.
U.-L.~P.~acknowledges support from the Natural Sciences and Engineering Research Council of Canada.
Research at Perimeter Institute is supported by the Government of Canada through the Department of Innovation, Science and Economic Development Canada  and by the Province of Ontario through the Ministry of Research, Innovation and Science.
J.~B.~P.~acknowledges support from NSF Award 1211777. 
Computations were performed on the GPC supercomputer at the SciNet HPC Consortium.

\medskip

\bibliographystyle{mnras}
\bibliography{references.bib}

\begin{thebibliography}{}
\makeatletter
\relax
\def\mn@urlcharsother{\let\do\@makeother \do\$\do\&\do\#\do\^\do\_\do\%\do\~}
\def\mn@doi{\begingroup\mn@urlcharsother \@ifnextchar [ {\mn@doi@}
  {\mn@doi@[]}}
\def\mn@doi@[#1]#2{\def\@tempa{#1}\ifx\@tempa\@empty \href
  {http://dx.doi.org/#2} {doi:#2}\else \href {http://dx.doi.org/#2} {#1}\fi
  \endgroup}
\def\mn@eprint#1#2{\mn@eprint@#1:#2::\@nil}
\def\mn@eprint@arXiv#1{\href {http://arxiv.org/abs/#1} {{\tt arXiv:#1}}}
\def\mn@eprint@dblp#1{\href {http://dblp.uni-trier.de/rec/bibtex/#1.xml}
  {dblp:#1}}
\def\mn@eprint@#1:#2:#3:#4\@nil{\def\@tempa {#1}\def\@tempb {#2}\def\@tempc
  {#3}\ifx \@tempc \@empty \let \@tempc \@tempb \let \@tempb \@tempa \fi \ifx
  \@tempb \@empty \def\@tempb {arXiv}\fi \@ifundefined
  {mn@eprint@\@tempb}{\@tempb:\@tempc}{\expandafter \expandafter \csname
  mn@eprint@\@tempb\endcsname \expandafter{\@tempc}}}

\bibitem[\protect\citeauthoryear{Anderson}{Anderson}{1993}]{Anderson:1993cna}
Anderson S.~B.,  1993, PhD thesis, Caltech, \url
  {http://resolver.caltech.edu/CaltechETD:etd-08282008-094151}

\bibitem[\protect\citeauthoryear{{Bandura} et~al.,}{{Bandura}
  et~al.}{2014}]{2014SPIE.9145E..22B}
{Bandura} K.,  et~al., 2014, in Ground-based and Airborne Telescopes V. p.
  914522 (\mn@eprint {arXiv} {1406.2288}), \mn@doi{10.1117/12.2054950}

\bibitem[\protect\citeauthoryear{{Camilo}, {Nice}, {Shrauner}  \&
  {Taylor}}{{Camilo} et~al.}{1996}]{1996ApJ...469..819C}
{Camilo} F.,  {Nice} D.~J.,  {Shrauner} J.~A.,   {Taylor} J.~H.,  1996, \mn@doi
  [\apj] {10.1086/177829}, \href
  {http://adsabs.harvard.edu/abs/1996ApJ...469..819C} {469, 819}

\bibitem[\protect\citeauthoryear{{Chang}, {Pen}, {Bandura}  \&
  {Peterson}}{{Chang} et~al.}{2010}]{2010Natur.466..463C}
{Chang} T.-C.,  {Pen} U.-L.,  {Bandura} K.,   {Peterson} J.~B.,  2010, \mn@doi
  [\nat] {10.1038/nature09187}, \href
  {http://adsabs.harvard.edu/abs/2010Natur.466..463C} {466, 463}

\bibitem[\protect\citeauthoryear{{Cordes} \& {Lazio}}{{Cordes} \&
  {Lazio}}{2002}]{2002astro.ph..7156C}
{Cordes} J.~M.,  {Lazio} T.~J.~W.,  2002, preprint, \href
  {http://adsabs.harvard.edu/abs/2002astro.ph..7156C} {} (\mn@eprint {}
  {astro-ph/0207156})

\bibitem[\protect\citeauthoryear{{DeBoer} et~al.,}{{DeBoer}
  et~al.}{2017}]{2017PASP..129d5001D}
{DeBoer} D.~R.,  et~al., 2017, \mn@doi [\pasp]
  {10.1088/1538-3873/129/974/045001}, \href
  {http://adsabs.harvard.edu/abs/2017PASP..129d5001D} {129, 045001}

\bibitem[\protect\citeauthoryear{{Fonseca}, {Maartens}  \& {Santos}}{{Fonseca}
  et~al.}{2017}]{2017MNRAS.466.2780F}
{Fonseca} J.,  {Maartens} R.,   {Santos} M.~G.,  2017, \mn@doi [\mnras]
  {10.1093/mnras/stw3248}, \href
  {http://adsabs.harvard.edu/abs/2017MNRAS.466.2780F} {466, 2780}

\bibitem[\protect\citeauthoryear{{Glazebrook} et~al.,}{{Glazebrook}
  et~al.}{2007}]{2007ASPC..379...72G}
{Glazebrook} K.,  et~al., 2007, in {Metcalfe} N.,  {Shanks} T.,  eds,
  Astronomical Society of the Pacific Conference Series Vol. 379, Cosmic
  Frontiers. p.~72 (\mn@eprint {} {astro-ph/0701876})

\bibitem[\protect\citeauthoryear{{Hotan}, {van Straten}  \&
  {Manchester}}{{Hotan} et~al.}{2004}]{2004PASA...21..302H}
{Hotan} A.~W.,  {van Straten} W.,   {Manchester} R.~N.,  2004, \mn@doi [\pasa]
  {10.1071/AS04022}, \href {http://adsabs.harvard.edu/abs/2004PASA...21..302H}
  {21, 302}

\bibitem[\protect\citeauthoryear{{Johnston} \& {Karastergiou}}{{Johnston} \&
  {Karastergiou}}{2017}]{2017MNRAS.467.3493J}
{Johnston} S.,  {Karastergiou} A.,  2017, \mn@doi [\mnras]
  {10.1093/mnras/stx377}, \href
  {http://adsabs.harvard.edu/abs/2017MNRAS.467.3493J} {467, 3493}

\bibitem[\protect\citeauthoryear{{Lentati} \& {Shannon}}{{Lentati} \&
  {Shannon}}{2015}]{2015MNRAS.454.1058L}
{Lentati} L.,  {Shannon} R.~M.,  2015, \mn@doi [\mnras]
  {10.1093/mnras/stv2089}, \href
  {http://adsabs.harvard.edu/abs/2015MNRAS.454.1058L} {454, 1058}

\bibitem[\protect\citeauthoryear{{Lentati}, {Alexander}  \& {Hobson}}{{Lentati}
  et~al.}{2015}]{2015MNRAS.447.2159L}
{Lentati} L.,  {Alexander} P.,   {Hobson} M.~P.,  2015, \mn@doi [\mnras]
  {10.1093/mnras/stu2611}, \href
  {http://adsabs.harvard.edu/abs/2015MNRAS.447.2159L} {447, 2159}

\bibitem[\protect\citeauthoryear{{Lentati} et~al.,}{{Lentati}
  et~al.}{2017}]{2017MNRAS.466.3706L}
{Lentati} L.,  et~al., 2017, \mn@doi [\mnras] {10.1093/mnras/stw3359}, \href
  {http://adsabs.harvard.edu/abs/2017MNRAS.466.3706L} {466, 3706}

\bibitem[\protect\citeauthoryear{{Lorimer} \& {Kramer}}{{Lorimer} \&
  {Kramer}}{2004}]{2004hpa..book.....L}
{Lorimer} D.~R.,  {Kramer} M.,  2004, {Handbook of Pulsar Astronomy}

\bibitem[\protect\citeauthoryear{{Manchester}, {Hobbs}, {Teoh}  \&
  {Hobbs}}{{Manchester} et~al.}{2005}]{2005AJ....129.1993M}
{Manchester} R.~N.,  {Hobbs} G.~B.,  {Teoh} A.,   {Hobbs} M.,  2005, \mn@doi
  [\aj] {10.1086/428488}, \href
  {http://adsabs.harvard.edu/abs/2005AJ....129.1993M} {129, 1993}

\bibitem[\protect\citeauthoryear{{Masui} et~al.,}{{Masui}
  et~al.}{2013}]{2013ApJ...763L..20M}
{Masui} K.~W.,  et~al., 2013, \mn@doi [\apjl] {10.1088/2041-8205/763/1/L20},
  \href {http://adsabs.harvard.edu/abs/2013ApJ...763L..20M} {763, L20}

\bibitem[\protect\citeauthoryear{{Masui} et~al.,}{{Masui}
  et~al.}{2015}]{FRB110523}
{Masui} K.,  et~al., 2015, \mn@doi [\nat] {10.1038/nature15769}, \href
  {http://adsabs.harvard.edu/abs/2015Natur.528..523M} {528, 523}

\bibitem[\protect\citeauthoryear{{Newburgh} et~al.,}{{Newburgh}
  et~al.}{2016}]{2016SPIE.9906E..5XN}
{Newburgh} L.~B.,  et~al., 2016, in Ground-based and Airborne Telescopes VI. p.
  99065X (\mn@eprint {arXiv} {1607.02059}), \mn@doi{10.1117/12.2234286}

\bibitem[\protect\citeauthoryear{{Os{\l}owski}, {van Straten}, {Hobbs},
  {Bailes}  \& {Demorest}}{{Os{\l}owski} et~al.}{2011}]{2011MNRAS.418.1258O}
{Os{\l}owski} S.,  {van Straten} W.,  {Hobbs} G.~B.,  {Bailes} M.,   {Demorest}
  P.,  2011, \mn@doi [\mnras] {10.1111/j.1365-2966.2011.19578.x}, \href
  {http://adsabs.harvard.edu/abs/2011MNRAS.418.1258O} {418, 1258}

\bibitem[\protect\citeauthoryear{{Pennucci}}{{Pennucci}}{2013}]{2013PhDT..Pennucci}
{Pennucci} T.~T.,  2013, PhD thesis, University of Virginia (USA)

\bibitem[\protect\citeauthoryear{{Santos} et~al.,}{{Santos}
  et~al.}{2015}]{2015aska.confE..19S}
{Santos} M.,  et~al., 2015, Advancing Astrophysics with the Square Kilometre
  Array (AASKA14), \href {http://adsabs.harvard.edu/abs/2015aska.confE..19S}
  {p.~19}

\bibitem[\protect\citeauthoryear{{Smith}}{{Smith}}{2016}]{2016arXiv161006831S}
{Smith} K.~M.,  2016, preprint, \href
  {http://adsabs.harvard.edu/abs/2016arXiv161006831S} {} (\mn@eprint {arXiv}
  {1610.06831})

\bibitem[\protect\citeauthoryear{{Switzer} et~al.,}{{Switzer}
  et~al.}{2013}]{2013MNRAS.434L..46S}
{Switzer} E.~R.,  et~al., 2013, \mn@doi [\mnras] {10.1093/mnrasl/slt074}, \href
  {http://adsabs.harvard.edu/abs/2013MNRAS.434L..46S} {434, L46}

\bibitem[\protect\citeauthoryear{{Taylor}}{{Taylor}}{1992}]{1992RSPTA.341..117T}
{Taylor} J.~H.,  1992, \mn@doi [Philosophical Transactions of the Royal Society
  of London Series A] {10.1098/rsta.1992.0088}, \href
  {http://adsabs.harvard.edu/abs/1992RSPTA.341..117T} {341, 117}

\bibitem[\protect\citeauthoryear{{Yusifov} \& {K{\"u}{\c c}{\"u}k}}{{Yusifov}
  \& {K{\"u}{\c c}{\"u}k}}{2004}]{2004A&A...422..545Y}
{Yusifov} I.,  {K{\"u}{\c c}{\"u}k} I.,  2004, \mn@doi [\aap]
  {10.1051/0004-6361:20040152}, \href
  {http://adsabs.harvard.edu/abs/2004A%26A...422..545Y} {422, 545}

\makeatother
\end{thebibliography}

\bsp	
\label{lastpage}
\end{document}